\documentclass[%
 reprint,
 amsmath,amssymb,
 aps,
aps,prx,
]{revtex4-2}

\usepackage{graphicx} % Include figure files
\usepackage{dcolumn}  % Align table columns on decimal point
\usepackage{bm}       % bold math
\usepackage{verbatim} %add comments
\usepackage{gensymb}
\usepackage{lineno}   % numbers of rows
\usepackage{soul}     %
\usepackage{color}
\usepackage{textcomp}
\usepackage{CJK}

\begin{document}
%\nolinenumbers

\preprint{APS/123-QED}

\title{3D trapping and dynamic axial manipulation with frequency-tuned spiraling acoustical tweezers }% Force line breaks with \\
%Static holographic acoustical tweezers for dynamic manipulation: Focused vortex

\author{Zhixiong Gong}
\affiliation{Univ. Lille, CNRS, Centrale Lille, Univ. Polytechnique Hauts-de-France, UMR 8520 -
IEMN - Institut d’Électronique de Microélectronique et de Nanotechnologie, F-59000
Lille, France}%
\author{Michael Baudoin}%
\email{Corresponding author: michael.baudoin@univ-lille.fr}
\homepage{\mbox{http://films-lab.univ-lille1.fr/michael}}
\affiliation{Univ. Lille, CNRS, Centrale Lille, Univ. Polytechnique Hauts-de-France, UMR 8520 -
IEMN - Institut d’Électronique de Microélectronique et de Nanotechnologie, F-59000
Lille, France}%
\affiliation{Institut Universitaire de France, 1 rue Descartes, 75005 Paris}%

\date{\today}% It is always \today, today,

\begin{abstract}
Holographic acoustical tweezers (HAT) based on Archimedes-Fermat spiraling InterDigitated Transducers (S-IDTs)  are a versatile tool for the selective manipulation of microparticles [Baudoin \textit{et. al.}, Sci. Adv., 5: eaav1967 (2019)] and cells [Baudoin \textit{et. al.}, Nat. Commu., 11, 4244 (2020)] in a standard microfluidic environment. These binary active holograms produce some focused helical wave, with the ability to trap particles at the vortex core. Yet, all the studies conducted with S-IDTs have so far been restricted to 2D manipulation only. Here we show (i) that 3D radiation trap for microparticles and cells can be obtained with spiraling tweezers with sufficiently large aperture and (ii) that the particles can be displaced axially by simply tuning the driving frequency, without any motion of the transducer. This work opens perspectives for 3D cells and microparticles manipulation with single-beam acoustical tweezers.
\end{abstract}

\pacs{Valid PACS appear here}% PACS, the Physics and Astronomy
%\keywords{Suggested keywords}%Use showkeys class option if keyword display desired
\maketitle

%------------------------------------------------------------------------------------------------
\section{\label{sec:Introduction}Introduction}

The idea of manipulating objects with the acoustic radiation force was instilled by early theoretical work of Rayleigh \cite{pm_rayleigh_1902,pm_rayleigh_1905}, Langevin \cite{ra_biquard_1932a,ra_biquard_1932b} and Brillouin \cite{ap_brillouin_1925,jpr_brillouin_1925} on the acoustic radiation pressure and the first calculations of the force applied by an acoustic field on a spherical particle \cite{king1934acoustic,a_yosika_1955,spd_gorkov_1962}. Experimentally, some early demonstration of particles trapping at the nodes and anti-nodes of standing waves depending on their properties are reported at the beginning of the 20th century \cite{sp_boyle_1928,rsct_boyle_1929,jasa_allen_1947}. Since then, this principle has been refined, e.g. by combining high frequency surface acoustic wave orthogonal transducers and microfluidics techniques to manipulate microparticles and cells in 2D
in a microchannel \cite{apl_tran_2012,pnas_ding_2012}, or by using static \cite{meldenature2016} or reconfigurable holograms \cite{po_ochiai_2014,acmtg_ochiai_2014,marzo2019holographic,HirayamaVolumetric2019,nc_ma_2020} to pattern or manipulate many particles simultaneously in 2D and in 3D. In acoustics, the term "Acoustical tweezers" was first introduced by Wu \cite{wu1991acoustical} in analogy with optical tweezers. But in this work, like in all the aforementioned work, the acoustical tweezers lack two essential features of their optical counterpart \cite{ol_ashkhin_1986}: (i) \textit{the selectivity}, i.e. the ability to select, trap and manipulate a single object without affecting other neighboring objects and (ii) \textit{3D trapping capability with a single beam}, i.e. the ability to produce a 3D trap from a beam produced from only one side of the manipulation chamber. 

The selectivity requires spatial localization of the acoustic energy to only affect the targeted object \cite{arfm_baudoin_2020}. This localization can be achieved with focused beams as first investigated in optics \cite{ol_ashkhin_1986} and later on in acoustics \cite{apl_shung_2009}. But in acoustics, many particles of interest (in particular the ones with positive contrast factors) are expelled, not trapped, at the center of a focused beam, or can be trapped only laterally at very specific frequencies close to the particle resonances \cite{jasa_gong_2021}. To trap this type of particles, a possibility is to use acoustical vortices \cite{arfm_baudoin_2020} (Bessel beams of topological order larger than or equal to one), some helical wavefields spinning around a phase singularity, first investigated in acoustics by Hefner and Marston \cite{jasa_hefner_1999}. 
At first, the potential of cylindrical acoustical vortices to trap particle in 2D has been investigated both theoretically \cite{jasa_baresch_2013} and experimentally \cite{apl_courtney_2013}. Such wavefields can be produced with an array of bulk \cite{prl_thomas_2003,apl_courtney_2013,yang2021Resaerch} or SAW transducers \cite{prap_riaud_2015,pre_riaud_2015,ieee_riaud_2016}, passive holograms based on diffraction gratings \cite{pre_jimenez_2016,apl_wang_2016,apl_wang_2016,apl_jiang_2016}, phase plates \cite{mupb_terzy_2017} and metameterials \cite{prl_jiang_2016}, or finally active holograms based on spiraling interdigitated transducers (S-IDTs) \cite{prap_riaud_2017}. Transducers arrays have the advantage to produce reconfigurable acoustic fields, which can be controlled electronically to move the object. As a counterpart these systems require a complex array of transducer, which can be complicated to manufacture at ultrasonic high frequencies due to cross coupling and high-end costly electronics, which both severely limit the applicability of this method to trap microscopic particle. Passive holograms based on phased lens and diffraction gratings turn the signal produced by a single plane wave transducer into the desired wavefield. The advantage is the simplicity of the system. But due to the diffraction limit, the phased lens or diffraction grating engineering precision $d$ should not be larger than half of the wavelength $d \leq \lambda / 2$ \cite{meldenature2016}, which severely limits the frequency range that can be achieved. The S-IDTs on the other hand rely on spiraling electrodes patterned at the surface of a transparent piezoelectric substrate with classic photolithography techniques. Hence, they are flat, transparent, easily integrable  in a microscope \cite{prap_riaud_2017} and should enable vortex synthesis up to GHz frequency \cite{iee_brizoual_2008}.

With tweezers based on cylindrical vortices, (i) the selectivity remains limited \cite{arfm_baudoin_2020} since the beam is only focused laterally leading to spurious secondary rings whose intensity decrease as $1/r^2$ (ii) particles can only be pushed/pulled \cite{marston2006axial,zhang2011geometrical,prap_fan_2019,gong2019PRE} but not trapped in the axial direction, owing to the invariance of the intensity profile in this direction. Three-dimensional tweezers require that the particle is subjected to restoring acoustic radiation force from all directions at the same time. These two problems (reduced selectivity and 3D trapping inability) can be addressed by using focused acoustical vortices or spherical vortices as first demonstrated theoretically \cite{baresch2013spherical} and later on experimentally \cite{baresch2016observation}  by Baresch et al. in water. Levitation with vortex beams was also demonstrated in air by Marzo et al. \cite{nc_marzo_2015}, who also investigated the possibility offered by other types of beam such as bottle and twin beams to trap particles laterally and axially respectively. Applicability of these systems for in-vivo manipulation of millimeter-size objects in the urinary bladders of live pigs has recently been demonstrated \cite{pnas_bailey_2020}.

Nevertheless these systems based on transducers arrays suffer the same limitations as the one previously mentioned in the 2D case for micrometric objects in a standard microscopy environment. Jimenez et al. \cite{apl_jimenez_2018} and Baudoin et al. \cite{baudoin2019folding,baudoin2020spatially} demonstrated that focused vortices can be produced with passive Fresnel spiral zone plates and active spiraling IDTs (S-IDTs) with Archimedes-Fermat geometry respectively. In particular, Baudoin et al. \cite{baudoin2019folding,baudoin2020spatially} were able to synthesize high frequency acoustical vortices and demonstrate selective trapping and displacement of microparticles \cite{baudoin2019folding} and cells \cite{baudoin2020spatially} in a standard microscopy environment, with forces several orders larger than their optical counterparts \cite{baudoin2020spatially}. Nevertheless, the 3D trapping capabilities of these S-IDTs based tweezers were not investigated since the experiments were conducted in narrow microfluidic chambers with reduced axial dimension. Also, one difficulty with these integrated systems is that the transducers cannot be displaced axially to keep the contact with the microfluidic chamber, hence preventing axial manipulation of the particles. Recently, it was shown in airborne experiments \cite{Karen2020active} that the focal point of a focused vortex synthesized with a spiralling structure can be moved axially by tuning the activation frequency, while also leading to some slight distorsion of the acoustical vortex.

In this paper, we show theoretically (i) that a 3D radiation trap for microparticles and cells can be produced with single beam S-IDTs tweezers and (ii) that the 3D trap is maintained and displaced axially when tuning the driving frequency.

%-----------------------------------------------------------------------------------------------
\section{Method: angular spectrum based calculation of the 3D acoustic radiation force}

The 3D radiation force applied on a microparticle by an acoustic field produced by spiraling Archimedes-Fermat spiraling transducers is calculated with an in-house code \cite{baudoin2020spatially} based on the angular spectrum method (ASM) originally introduced by Sapozhnikov \& Bailey \cite{sapozhnikov2013radiation}. Outline of this method and comparison to other existing formula \cite{jasa_silva_2011,jasa_baresch_2013} of the 3D acoustic radiation force applied on a particle of arbitrary size is discussed in \cite{gong2020equivalence}.

The flowchart for using ASM to compute the wave propagation is given in Appendix \ref{sec:Appendix A} and discussed in details. Here, we will briefly review the ASM-based radiation force formulas: The incident field in a source plane is decomposed into a sum of plane waves using the angular spectrum decomposition (the 2D spatial Fourier Transform of the incident field). Then the known decomposition of plane waves into spherical harmonics and the Legendre addition theorem are used to express the incident field in the spherical harmonics basis weighted by the ASM-based beam shape coefficients $H_{nm}$:

\begin{equation}
p_{i}  =  \frac{1}{ \pi } \sum_{n=0}^{\infty} \sum_{m=-n}^{n} i^{n}  H_{n m} j_{n}(k r)  Y_{n }^{m}(\theta, \varphi) , 
\label{incident_pressure}
\end{equation}
 with 
 \begin{equation}
\begin{aligned}
H_{n m}=\iint_{k_{x}^{2}+k_{y}^{2} \leq k^{2}}  S\left(k_{x}, k_{y}\right) Y_{n }^{m*} \left(\theta_{k}, \varphi_{k}\right)d k_{x} d k_{y}
\end{aligned},
\label{Hnm}
\end{equation}
where $S\left(k_{x}, k_{y}\right)$ is the so-called angular spectrum, $j_{n}(k r)$ and $Y_{n}^{m}$ are, respectively, the spherical Bessel function of the first kind and the spherical harmonics, $k_x$ and $k_y$ are the wavevector ($\mathbf{k}$) components in $x$ and $y$ direction with the relations to angles $\cos\theta_k = [1-\left(k_{x}^{2}+k_{y}^{2}\right) / k^{2}]^{1/2}$ and $\varphi_k = \arctan \left(k_{y} / k_{x}\right)$.

\begin{figure*} [!htbp]
\includegraphics[width=17.6cm]{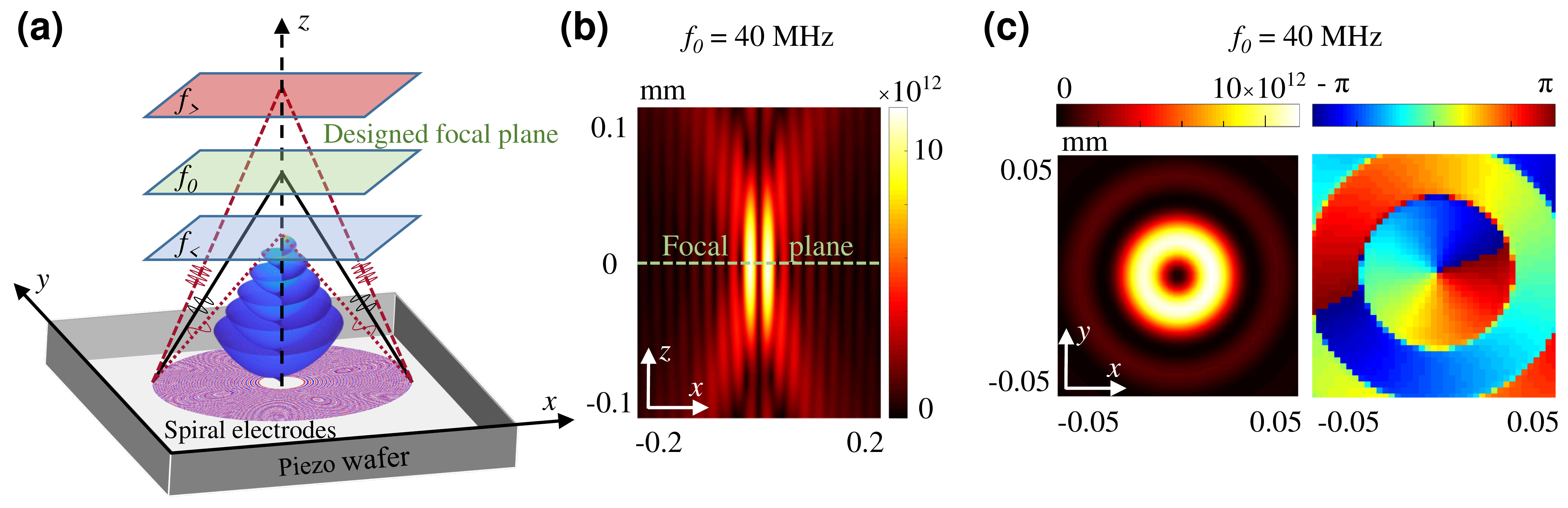}
\caption{Field produced by a one-sided holographic acoustical tweezers (HAT) designed to synthesize a focused vortex configuration. (a) The HAT is made of two inter-twinned spiralling electrodes of inverse polarity whose equations are given in appendix \ref{sec:Appendix B}. It is designed to synthesize a focus vortex at the original driving frequency $f_0=40$ MHz (wavelength $\lambda=$ 37.5 $\mu$m) and focus at height $z=0$, 1mm away from the source plane  $z=-1$ mm. The focal plane along the propagation direction can be changed by changing the actuation frequency. When the excitation frequency $f_>$ is higher than the designed frequency $f_0$, i.e., $f_> > f_0$, the focal plane moves away from the source plane, while when $f_< < f_0$, the focal plane moves closer to the source plane. (b) Pressure amplitude square in the propagation $(x,z)$ plane at actuation frequency $f_0$. (c) Pressure amplitude square and phase in the focal plane $(x,y), z=0$ at actuation frequency $f_0$.}
\label{Fig1: Schematic and acoustic field}
\end{figure*}

 The scattering problem is solved by using known results for planed waves  \cite{sapozhnikov2013radiation,gong2020acoustic} leading to:
\begin{equation}
p_{s} =  \frac{1}{\pi } \sum_{n=0}^{\infty} \sum_{m=-n}^{n} i^{n}  H_{n m} A_{n}^m h_{n}^{(1)}(k r) Y_{n}^{m}(\theta, \varphi) , \label{scattered_pressure}
\end{equation}
where $h_n^{(1)}$ is the spherical Hankel function of the first kind, and $A_{n}^m$ are the partial wave coefficients, which depend on the acoustic materials and geometric shape of the scatterer, and the boundary condition between the fluid and the particle. Finally, the force is obtained by integrating Brillouin's tensor \cite{ap_brillouin_1925} on a still surface in the far field  \cite{westervelt1951theory,westervelt1957acoustic} leading to the following equations of the 3D radiation force applied on a sphere in an arbitrary acoustic field obtained by Sapozhnikov \& Bailey \cite{sapozhnikov2013radiation} and slightly recast here under the form \cite{gong2020equivalence}:

\begin{widetext}  % write long formulas in two-column
\begin{subequations}
\begin{eqnarray}
F_{x}&=&\frac{1}{4 \pi^{2} \rho_{0} k^{2} c^{2}}  \operatorname{Re} \left\{ \sum_{n=0}^{\infty} \sum_{m=-n}^{n} C_{n} \left(-b_{n+1}^{-m} 
H_{n m} H_{n+1, m-1}^*
+ b_{n+1}^{m} H_{n m} H_{n+1, m+1}^*
 \right) \right\}, \label{ASM_Fx}
\\
F_{y}&=&\frac{1}{4 \pi^{2} \rho_{0} k^{2} c^{2}} \operatorname{Im} \left\{ \sum_{n=0}^{\infty} \sum_{m=-n}^{n} C_{n} b_{n+1}^{m} \left(  H_{n,-m} H_{n+1, -m-1}^* +  H_{n m} H_{n+1, m+1}^* 
\right) \right\}, \label{ASM_Fy}
\\
F_{z}&=&-\frac{1}{2 \pi^{2} \rho_{0} k^{2} c^{2}} \operatorname{Re} \left\{ \sum_{n=0}^{\infty} \sum_{m=-n}^{n} C_{n}^m c_{n+1}^{m} 
H_{n m} H_{n+1, m}^* \right\} . \label{ASM_Fz}
\end{eqnarray}
\label{MEM Torque}
\end{subequations}
\end{widetext}
where $C_n = A_{n}+2 A_{n} A_{n+1}^*+A_{n+1}^*$, $b_{n}^{m}=\sqrt{[{(n+m)(n+m+1)}]/[{(2 n-1)(2 n+1)}]}$ and $c_{n}^{m}=\sqrt{[{(n+m)(n-m)}]/[{(2 n-1)(2 n+1)}]}$. Note that here the partial wave coefficients $A_{n}^m$ reduce to $A_n$ owing to the spherical shape of the particle.

\section{\label{sec: 3D trapping}3D trapping with spiraling-IDT based tweezers}

%------------------------------------------------------------------------------------------------
\subsection{\label{sec: HAT design}HAT design and acoustic field}

Most particles of interest for microfluidic applications (e.g. solid microparticles) and biological applications (e.g. cells in water) are stiffer and denser than the surrounding medium and cannot be trapped in 3D with a one-sided focused beam \cite{jasa_gong_2021}. 3D selective trapping of such particles require specific wavefields such as focused acoustical vortices \cite{arfm_baudoin_2020}. Here we investigate the trapping capabilities of focused one-sided vortex beams resembling the fields synthesized by the spiraling IDTs used in references \cite{baudoin2019folding,baudoin2020spatially,gong2020acoustic}. It is produced by two Archimedes-Fermat spiraling electrodes of inverse polarity obtained by dicretizing the intersection of an acoustical vortex with a plane on two phase levels. The exact transducer geometry and source field used in the present simulations are  given in Appendix \ref{sec:Appendix B} and the resulting acoustic field is represented on Fig. \ref{Fig1: Schematic and acoustic field}. The transducer is designed to produce a focused vortex with an aperture of $65^{\circ}$ and a focal plane located 1mm away from the transducer when actuated at the original frequency $f_0 = 40$ MHz. The pressure amplitude square in the propagation plane ($x,y=0,z$) and the focal plane ($x,y,z=0$) are represented  on Fig. \ref{Fig1: Schematic and acoustic field} (b) and (c) respectively.

Now we will (i) investigate the ability of such acoustic field to trap a particle or a cell in 3D at the original frequency $f_0$ and (ii) investigate the evolution of the trap when the excitation frequency is slightly shifted away from $f_0$. 

\begin{figure} [!tbp]
\includegraphics[width=1\linewidth]{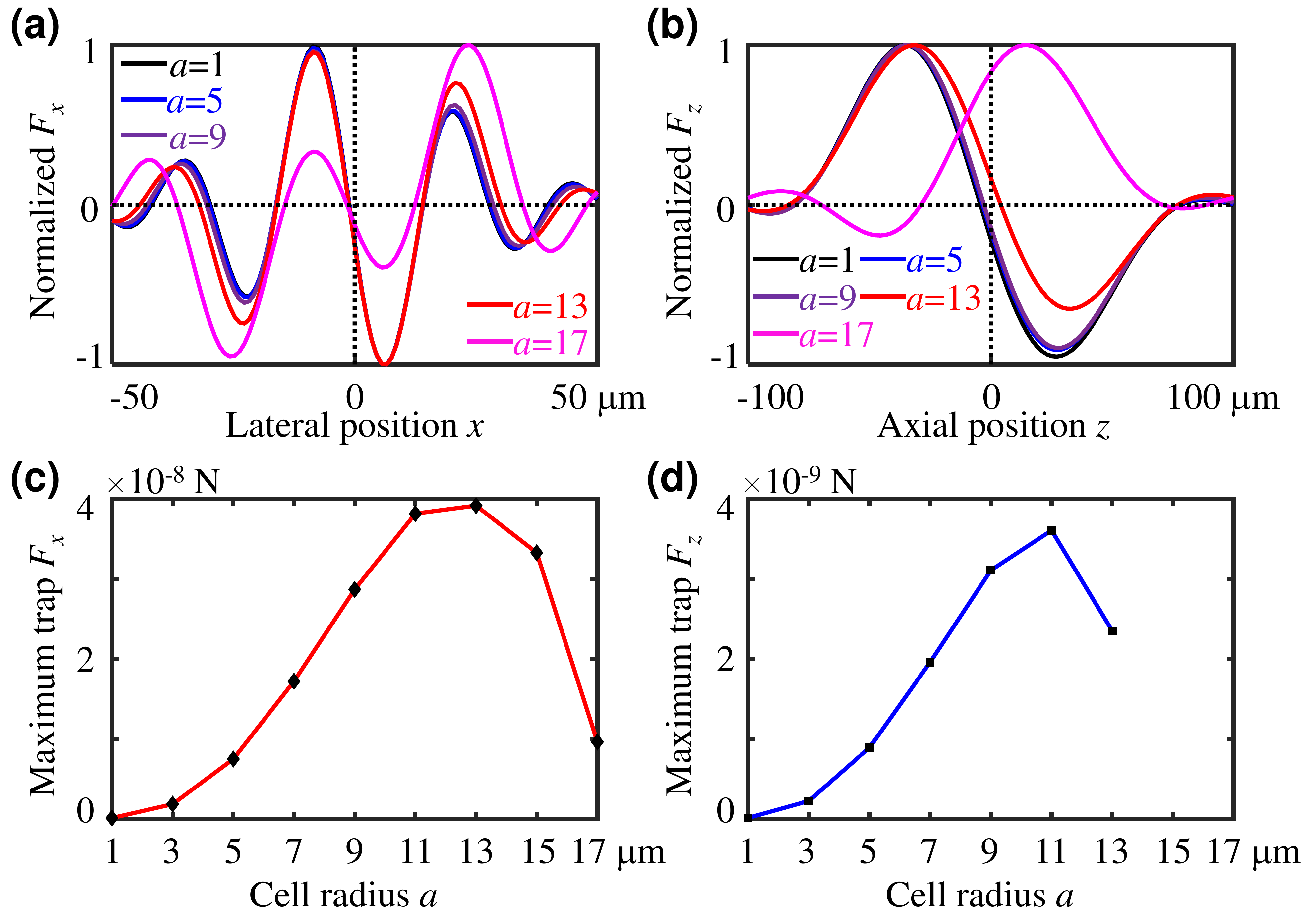}
\caption{Three-dimensional (3D) cell trapping with a spiraling one-sided Holographic Acoustic Tweezers (HAT) driven at the frequency $f_0=40$ MHz. The cell is modeled as a liquid sphere with typical acoustical parameters given in Table \ref{table1}. (a) and (b) represent the normalized lateral force vs $x$ position (at axial equilibrium position) and normalized axial force vs $z$ position respectively for cells with different radii $a$. 
The radius of the first ring of maximum intensity of the acoustical vortex is 13.6 $\mu$m (see Fig. \ref{Fig1: Schematic and acoustic field}.c), corresponding to 0.36$\lambda$.
(c) and (d) represent the maximum lateral and axial trapping (restoring) force for cells with radii ranging from $a=1$ to 17 $\mu$m. Only the stable trapping sizes are represented. The axial trap is lost for cell radius $a=$ 15 and 17 $\mu$m.}
\label{Fig2: 3D cell trap for different radii.}
\end{figure}

\begin{table*}[!htbp]
\small
  \caption{ Acoustic properties}
  \label{table1}
  \begin{tabular*}{1\textwidth}{@{\extracolsep{\fill}}lcccc}
    \hline 
    Material & Density $\rho_0$ ($kg/m^3$) & Longitudinal speed of sound $c$ (m/s) & Shear speed of sound $c$ (m/s) \\
    \hline 
    Cell       & 1100   & 1508   & ...            \\
    Pyrex      & 2230   & 5640   & 3280           \\
    Water    & 1000  & 1500   & ...         \\
    \hline
  \end{tabular*}
\end{table*}

%------------------------------------------------------------------------------------------------
\subsection{\label{sec: 3D cell trapping} 3D cell trapping}
\begin{figure} [!htbp]
\includegraphics[width=8.6cm]{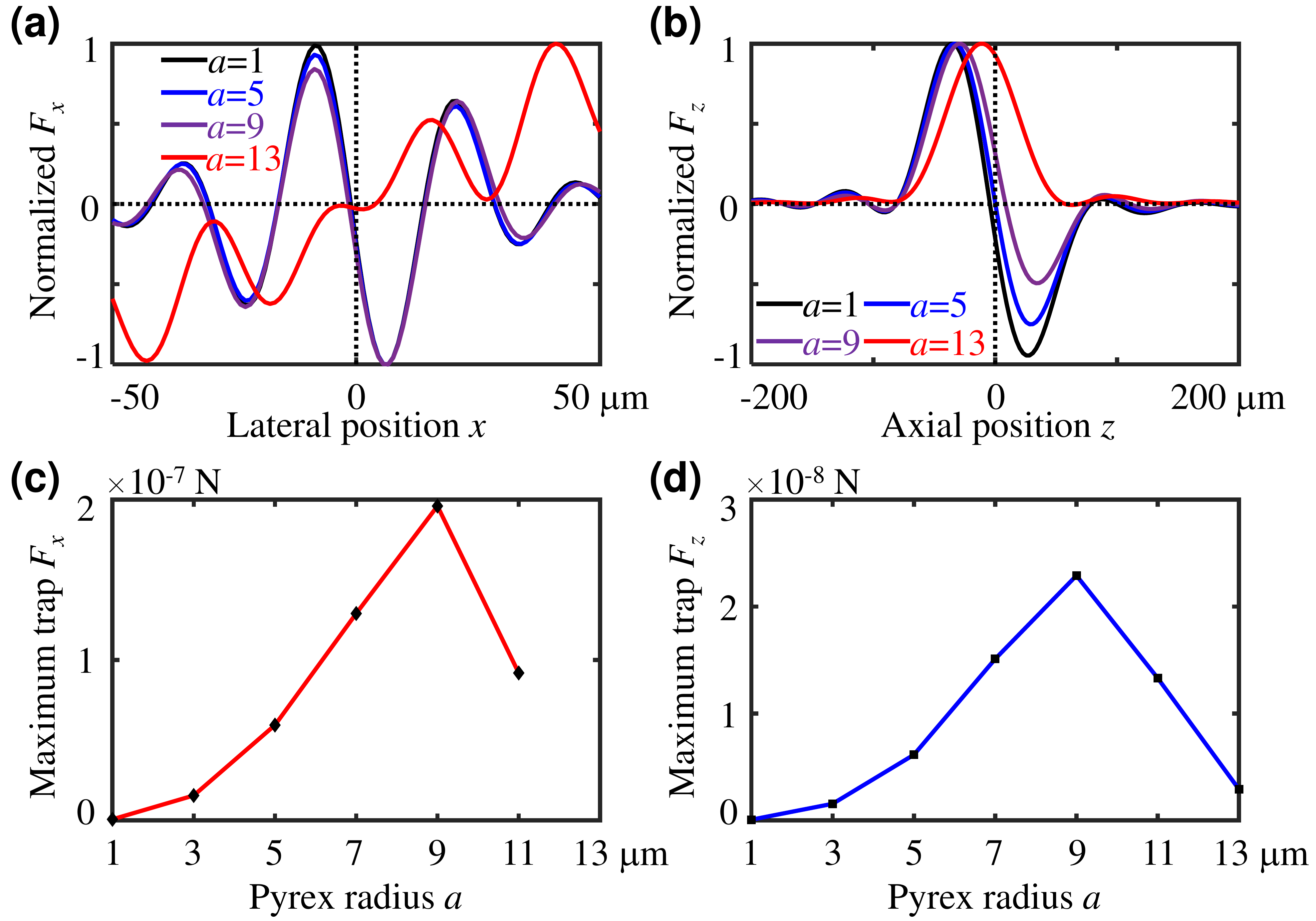}
\caption{3D Pyrex microparticle trapping with a vortex-based one-sided HAT at the designed frequency $f_0=40$ MHz with different radii $a$. The Pyrex is model as an elastic sphere without consideration of viscosity. The acoustical parameters are given in Table \ref{table1}. 
(a) and (b) are the normalized lateral force vs $x$ (at the axial equilibrium position) and normalized axial force vs $z$ position, respectively, for particle radii $a=1$, 5, 9, and 13 $\mu$m. 
(c) and (d) are the maximum trapping (restoring) force for particle radii ranging from $a=1$ to 13 $\mu$m. The lateral trapping is lost for Pyrex radius $a=$ 13 $\mu$m at the axial equilibrium position $z_{trap}=$ 57.5 $\mu$m.}
\label{Fig3: Radiation force on Pyrex.}
\end{figure}

\begin{figure*} [!htbp]
\includegraphics[width=17.6cm]{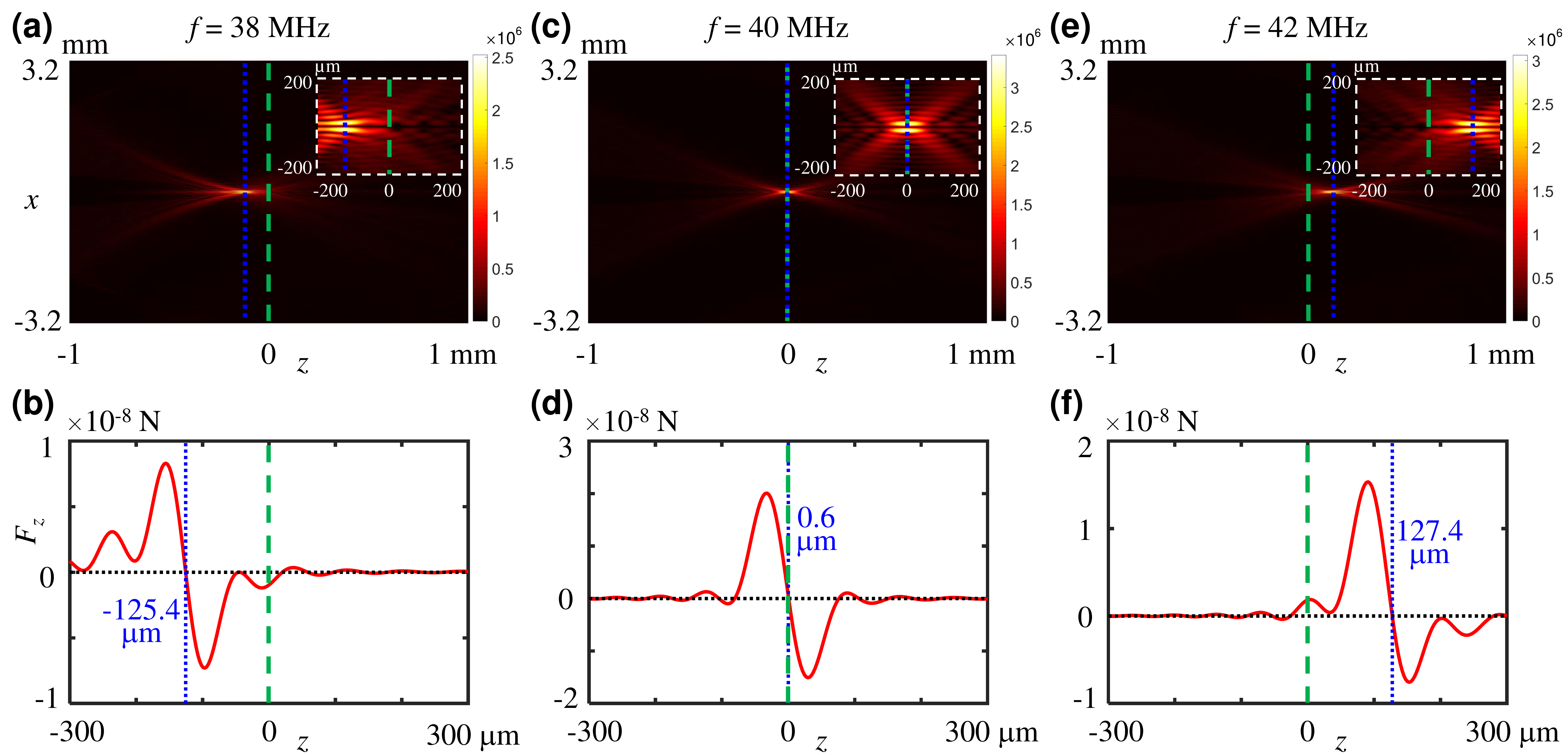}
\caption{
(a, c, e) The acoustic pressure amplitude in the propagation plane $(x,z)$ with different activated frequencies $f=$ 38, 40 and 42 MHz, respectively. The zoom in insets including the focal plane are shown in the top right corner.
(b, d, f) The axial radiation force versus different $z$ position for a Pyrex particle with radius $a=$ 7 $\mu$m. The Pyrex particle can be trapped in 3D (the lateral force versus $x$ position is not shown here for brevity) and the axial equilibrium position is changed with different activated frequency. The green dashed line describes the originally designed axial focal position ($z=0$) at $f_0=$ 40 MHz, while the blue dotted line indicates the actual axial position of the focal plane. }
\label{Fig4: acoustic field and trapping at different frequencies.}
\end{figure*}

Cells 2D trapping capability of S-IDTs has been recently verified both theoretically and experimentally  \cite{baudoin2020spatially}. Here we investigate the 3D trapping capabilities. The simulations are performed for a cell immersed in water and insonified by an acoustical vortex synthesized at the driving frequency $f_0= 40$ MHz by a binary spiralling hologram. The maximum pressure amplitude in the focal plane ($z=0$ mm) is 3.4 MPa. The cell is assimilated to a spherical liquid scatterer whose acoustical properties are provided in TABLE \ref{table1} and correspond to typical values provided in ref. \cite{augustsson2016iso}.

The computed normalized lateral ARF versus $x$ and axial ARF versus $z$ for a cell with different radii $a=$ 1, 5, 9, 13, and 17 $\mu$m are calculated and represented on Fig. \ref{Fig2: 3D cell trap for different radii.} (a) and (b), respectively. Note that the lateral force is computed at the axial trap equilibrium position $z_{trap}$. In addition, the maximum trapping force versus cell radii $a$ in the lateral and axial directions are represented on Fig. (c) and (d). These graph show that the S-IDTs based HAT produce a lateral trap for cells with radius up to $a=$ 17 $\mu$m (0.45$\lambda$), even larger than the radius of the maximum first ring  (0.36$\lambda$) in the lateral direction, while an axial trap is obtained for radii up to $a=$ 13 $\mu$m (0.35$\lambda$). That is to say, the designed HAT produce a 3D radiation trap of the considering typical cell with radius up to $min$ \{0.35$\lambda$,0.45$\lambda$\} $=$ 0.35$\lambda$. Note that in Fig. \ref{Fig2: 3D cell trap for different radii.}(b), the axial trapping position is shifted away from the source plane when the size of the cell is increased. Indeed, for small cells (see $a=$ 1, 5 and 9 $\mu$m), the gradient force is dominant and the scattering force is negligible \cite{gong2020three}. So the cells are trapped at or near the focal point. When the radius of the cell is increased, the scattering force can no more be neglected and will push the cell along the propagation direction. The restoring force remains nevertheless sufficient to balance the pushing force for cells smaller than $0.35 \, \lambda$, but the equilibrium position is shifted towards the beam direction of propagation (see $a=$ 13 and 17 $\mu$m). When the radius is further increased, the cell will escape from the trap. Note that the trapping forces calculated here for a realistic acoustic field are several orders of magnitude larger than the force resulting from the balance between gravity and buoyancy ($4.2 \times 10^{-15}$ Newton (N) for radius $a=$ 1 $\mu$m, and $9.2 \times 10^{-12}$ N for radius $a=$ 13 $\mu$m). 

%------------------------------------------------------------------------------------------------Sec III
\subsection{\label{sec:3D Pyrex trapping} 3D Pyrex trapping}
In this section, we investigate 3D trapping of spherical Pyrex particles with different radii from $a=1$ to $a=13$ $\mu$m in water with the same acoustic field as the one used in the previous section for cell trapping. Since Pyrex support both longitudinal and transverse waves, an elastic scattering model is considered \cite{baresch2016observation,gong2020three}. The acoustic properties of Pyrex used for the simulations are given in TABLE \ref{table1}.
The normalized lateral and axial radiation forces versus the respective positions are given in Fig. \ref{Fig3: Radiation force on Pyrex.}(a) and (b), with the maximum lateral and axial trapping (restoring) force versus particle radii $a$ in (c) and (d), respectively. These results show 3D Pyrex particle trapping for radii up to $a=$ 11 $\mu$m [0.29$\lambda$, see Fig. \ref{Fig3: Radiation force on Pyrex.}(c) and (d)], with a maximum for both lateral and axial trapping obtained around $a=$ 9 $\mu$m. It is interesting to note that the lateral trapping at the axial equilibrium position $z_{trap}$ is lost for $a=$ 13 $\mu$m since the axial trapping position is far away ( $z_{trap}=$ 57.5 $\mu$m) from the designed focus by the large pushing scattering force.
Similarly to cells, the gravity effect on a Pyrex sphere are negligible compared to the radiation force since the balance between gravity and buoyancy leads to net forces of  $5.2 \times 10^{-14}$ N for radius $a=$ 1 $\mu$m, and $6.9 \times 10^{-11}$ N for radius $a=$ 11 $\mu$m.

Note that the influence of the aperture on the 3D trapping capabilities of the tweezers for both Pyrex particles and cells are discussed in Appendix \ref{sec:Appendix C}.

%------------------------------------------------------------------------------------------------Sec 
\section{\label{sec: axial displacement} Axial displacement of the trap by frequency tuning}

\subsection{\label{sec:focaldisplacement} Axial displacement of the focal point.}

3D trapping ability of spherical focused vortices produced by HAT has been verified in Sec. \ref{sec: 3D trapping}. Now we will investigate what happens to the trapping equilibrium position when the actuation frequency of the HAT is tuned, while  keeping the geometric shapes of the IDTs identical. Indeed, recent airborne investigations \cite{Karen2020active} reported that slight frequency tuning of the excitation of a spiraling HAT, induce an axial translation of the position of the focal plane, combined with a distorsion of the beam since the equations of the Archimedes-Fermat spirals \cite{baudoin2019folding} are not invariant by a change in the tuned frequency. However they did not studied the trapping capabilities of these beams and whether the distortion remains sufficiently limited to maintain a 3D trap.

As illuminated in Fig. \ref{Fig1: Schematic and acoustic field}(a), the focal plane is moved toward the source plane (i.e., focal depth decreases) when the excitation frequency $f < f_0$ [see Fig. \ref{Fig4: acoustic field and trapping at different frequencies.}(a) for $f=38$ MHz], and away from it when $f > f_0$ [see Fig. \ref{Fig4: acoustic field and trapping at different frequencies.}(e) for $f = 42$ MHz]. 

\subsection{\label{sec:pyrexdisplacement} Axial displacement of Pyrex particles}

In this section, radiation forces simulations are conducted for a Pyrex spherical particles of radius $a=$ 7 $\mu$m, so that the lateral and axial trapping force are large and the restoring force around the equilibrium position is more symmetrical (compared to the case $a=$ 9 $\mu$m). 
The axial radiation force versus axial position $z$ near the designed focus ($z$ = 0) for the excitation frequency $f=$ 38, 40 and 42 MHz are represented on Fig. \ref{Fig4: acoustic field and trapping at different frequencies.}. (b), (d) and (f), respectively. The Pyrex sphere can be trapped axially at the new focal plane at $z_{trap}=-$125.4 $\mu$m for $f=$ 38 MHz, and  at $z_{trap}=$ 127.4 $\mu$m for $f=$ 42 MHz. The axial translation of the trapping point for these two cases is of the order of 18$a$ which is large enough for axial control of particle in a microchannel. Lateral trapping at the axial equilibrium position $z_{trap}$ (not shown here for the sake of brevity) is also fulfilled  since the main energy is localized near the focal plane with a large intensity gradient. Hence these simulations demonstrate controllable axial displacement of Pyrex particles trap when tuning the excitation frequency.

\begin{figure} [!htbp]
\includegraphics[width=8.6cm]{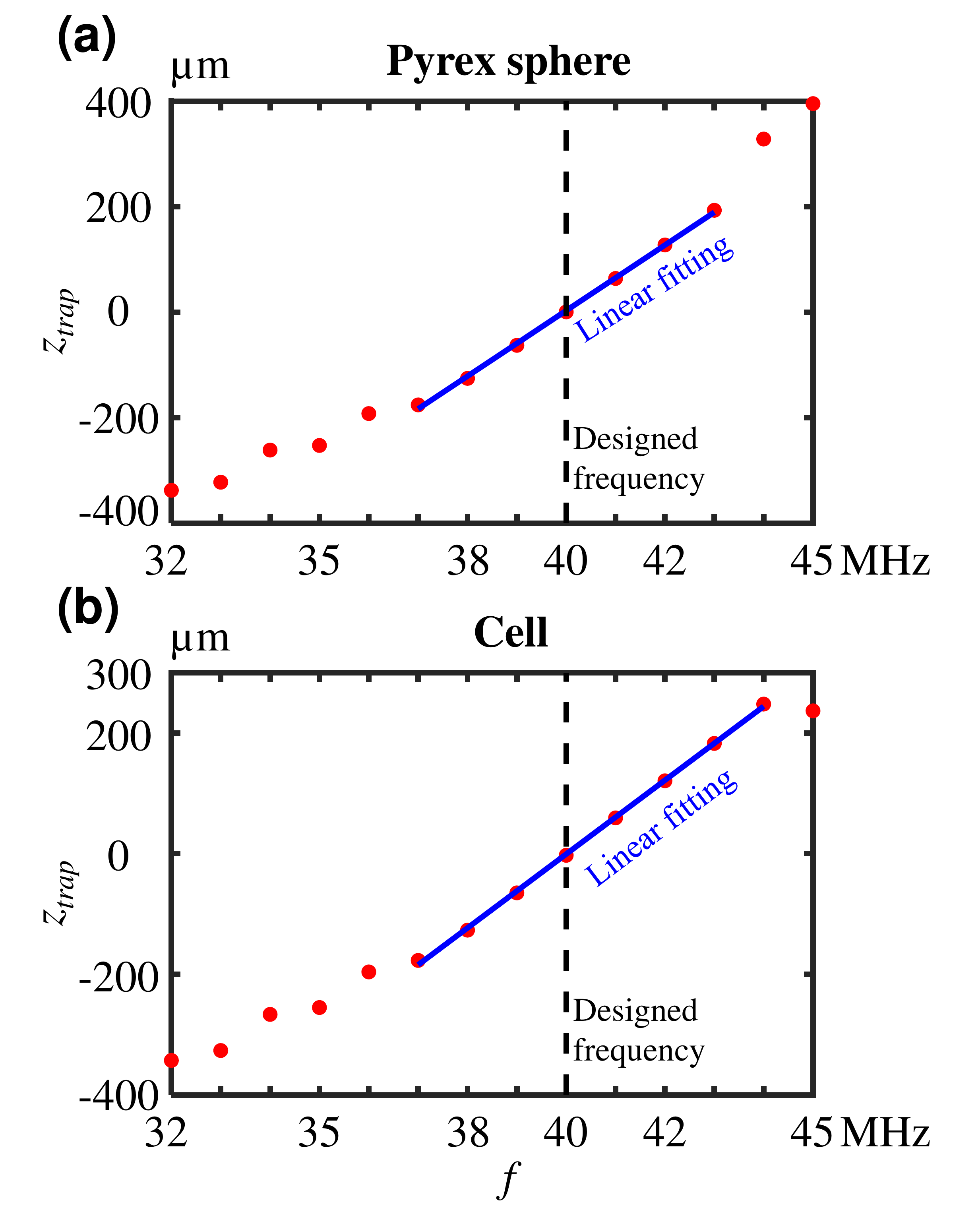}
\caption{The axial trapping positions along the $z$ direction versus excitation frequencies for (a)  Pyrex  spherical  particles with radius $a$ = 7 $\mu$m and (b) cells with $a$ = 11 $\mu$m. 
The excitation frequencies are selected near the designed frequency $f_0=$ 40 MHz from $f=$ 32 MHz to $f=$ 45 MHz.
In the regime of [37,43] MHz for (a) the Pyrex sphere, and of [37,44] MHz for (b) the cell, the axial trapping position $z_{trap}$ agrees exactly well with the linear fitting (blue solid line).}
\label{Fig5: z trapping position vs frequency.}
\end{figure}

We further conducted a systematic analysis of the axial translation of the trapping point versus the excitation frequency within the regime $f=$ [32,45] MHz [see Fig. \ref{Fig5: z trapping position vs frequency.}(a)]. When the excitation frequency is close to the designed frequency $f_0=$ 40 MHz, the trapping position $z_{trap}$ is linearly dependent on the excitation frequencies (see the regime  of [37,43] MHz). For excitation frequencies further away from the original frequency $f_0$, e.g., $f=$ [32,36] or [44,45] MHz, the trend is no more linear due to large distortion of the acoustic vortex. In the range investigated, the frequency shift enables to move the particle of $\sim 800 \mu$m, that is to say more than 100 times the size of the particle, and 20 times the wavelength.  

\subsection{\label{sec:celldisplacement} Axial displacement of cells.}

We conducted a similar analysis for cells with radius $a=$ 11 $\mu$m [see Fig. \ref{Fig5: z trapping position vs frequency.}(b)]. The analysis shows that cells can also be moved, while keeping a 3D trap when the frequency is tuned in the range $f=$ [32,45] MHz. This leads to a maximum displacement range of $\sim 650 \mu$m with a linear trend observed in the regime $f=$ [37,44] MHz, which is more than 57 times the size of the cell, and 16 times the wavelength.

%------------------------------------------------------------------------------------------------
\section{\label{sec:conclusion} Conclusion and discussion}
This work demonstrates theoretically the ability to obtain a 3D acoustic radiation traps for both elastic microparticles and cells with realistic acoustic fields mimicking the one produced with Spiraling Interdigitated Transducers (S-IDTs). This extends to  3D the manipulation possibilities offered by spiraling InterDigitated Transducers  \cite{baudoin2019folding,baudoin2020spatially,arfm_baudoin_2020}. In addition, we also demonstrate that the position of the 3D trapped cells/microparticles can be moved axially in a predictable manner by tuning the excitation frequency near the original frequency used for the design of the spiraling transducers. The magnitude of the displacement which can be achieved is two orders of magnitude larger than the particle radius and more than one order of magnitude the wavelength. This opens possibilities for axial displacement of the trapped particle by a simple shift of the driving frequency without any moving part. This is critical since for manipulation in microchannels, the tweezers cannot be moved axially. Next steps include (i) experimental verification of these predictions and (ii) investigation the role played by Eckart streaming \cite{pre_riaud_2014} and the resulting drag, especially for particles with low acoustic contrast such as cells.

%------------------------------------------------------------------------------------------------
\begin{acknowledgments}
We acknowledge the support from the ERC Generator, Prematuration programs, and Talent project funded by ISITE-ULNE and from Institut Universitaire de France.
\end{acknowledgments}

%------------------------------------------------------------------------------------------------
\appendix

%-------------------------------------
\section{\label{sec:Appendix A} Angular spectrum method}
\begin{figure} [!htbp]
\includegraphics[width=8.6cm]{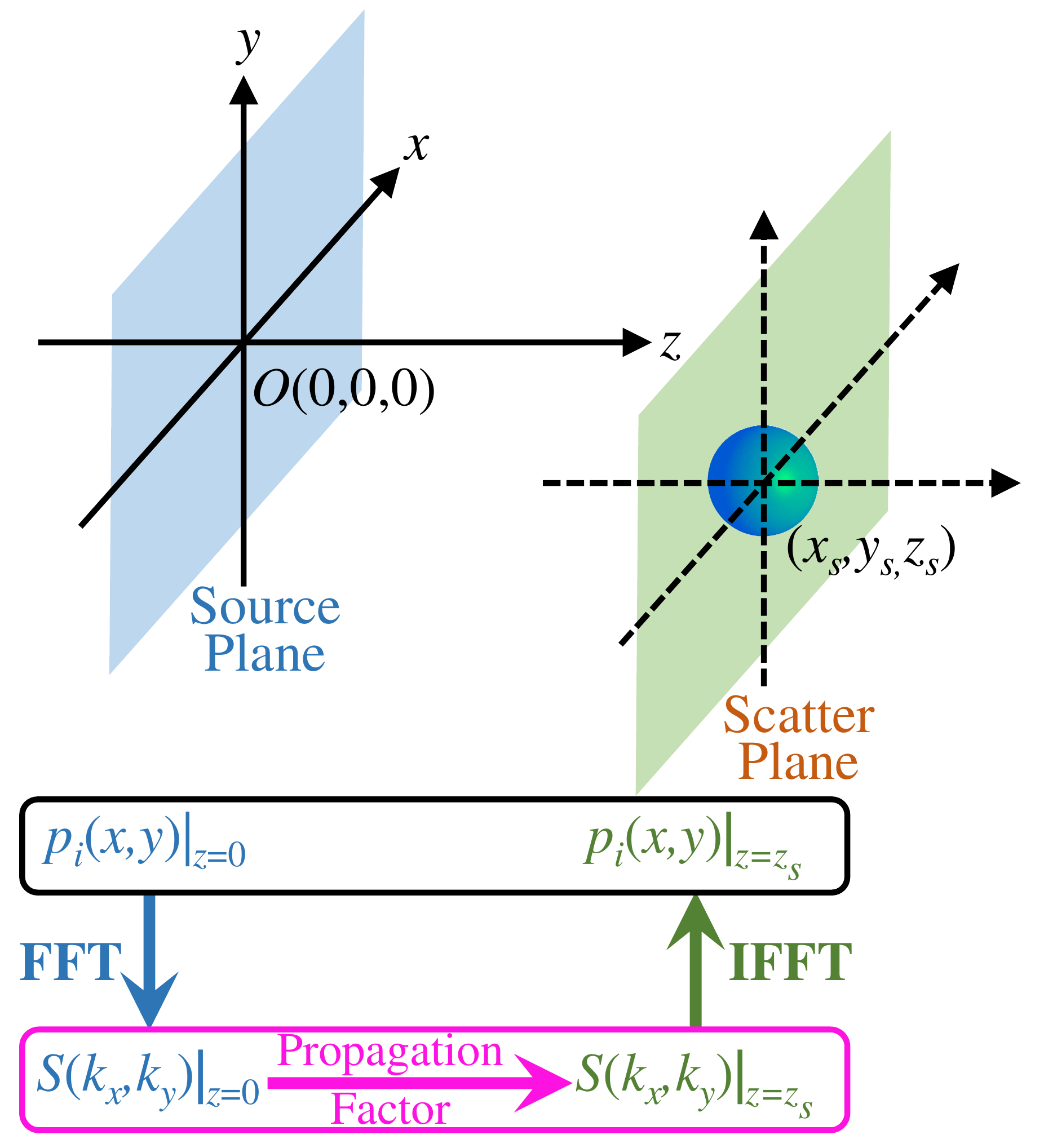}
\caption{The flowchart of angular spectrum method: The acoustic pressure in the black box lies in the real physical space and can be measured in experiment, while the the angular spectrum in the magenta box lies in the Fourier space.}
\label{Fig6: ASM Schematic}
\end{figure}
The schematic to understand how to use ASM to calculate the wave propagation in free space is given in Fig. \ref{Fig6: ASM Schematic}. Assume we know the acoustic pressure field in the source plane or a  plane parallel to the source plane, the angular spectrum $S(k_x,k_y)|_{z=0}$ (e.g., at source plane) is easily obtained by using the 2D Fast Fourier Transform (FFT) which is related to the ASM-based beam shape coefficients $H_{nm}$, see Eq. (\ref{Hnm}). 
To calculate the radiation force according to Eqs. (\ref{ASM_Fx}-\ref{ASM_Fz}), it is convenient to expand the incident beam at the origin of the scatter ($x_s,y_s,z_s$). The propagation of the acoustic field from the source to the scatter plane is calculated  by using the propagation factor in the Fourier space:
\begin{equation}
\begin{aligned}
S\left(k_{x}, k_{y}\right)|_{z=z_s} =& {S}\left(k_{x}, k_{y}\right)|_{z=0} \times \\
& e^{i k_{x} x_{s}+i k_{y} y_{s}+i \sqrt{k^{2}-k_{x}^{2}-k_{y}^{2}} z_{s}},
\label{eq:propagation factor}
\end{aligned}
\end{equation}
which can be used to calculate the ASM-based beam shape coefficients $H_{nm}$ at the scatterer center and further the 3D radiation force. Note also that the real pressure field including amplitude and phase at the scatter plane can be obtained by using the inverse 2D Fourier transform (IFFT) of the angular spectrum $S(k_x,k_y)|_{z=z_s}$.

%-------------------------------------
\section{\label{sec:Appendix B} Principle of focused vortex based acoustical tweezers}
It is difficult to produce a ideal spherical vortex in experiment since the transducers need to place all over the 4$\pi$ steradians enclosing the manipulation area \cite{gong2020three}. However, this can be overcome by using a focused spherical vortex with finite aperture (steradian less than 2$\pi$) \cite{baresch2013spherical,zhao2020generation,baudoin2019folding,baudoin2020spatially}.
Similar as the plane standing waves consisting of two counterpropagating travelling waves, a spherical Bessel beam can be produced by the interference of a converging and diverging spherical Hankel beams. This agrees with the mathematical fact that a Bessel function can be written as the addition of Hankel functions of the first and second kind (with a ratio of $1/2$).
To coincide with the radiation force derivations \cite{gong2019PRE,gong2020equivalence}, the time harmonics here is chosen as $e^{(-i \omega t)}$ so that the converging Hankel vortex is described by the spherical Hankel function of second kind $h_n^{(2)}$.
An ideal converging Hankel spherical vortex is described in the spherical coordinate $(r,\theta, \varphi)$ by \cite{baudoin2019folding}
\begin{equation}
p^{*}(r, \theta, \varphi) = A h_{n}^{(2)}(k r) P_{n}^{m}(\cos \theta) e^{i(m \varphi-\omega t)}
\label{eq:ideal Hankel vortex}
\end{equation}

Following the same procedure as in Ref. \cite{baudoin2019folding} and using the far-field asymptotic expression $h_{n}^{(2)}(k r) \simeq i^{(n+1)} e^{-ikr}/(kr)$, the geometrical shapes of the two spiral eletrodes can be described in cylindrical coordinates $(R, \varphi,z)$ as:

\begin{equation}
R_{1}=\frac{1}{k} \sqrt{\left(\varphi + C_{2}\right)^{2}-(k z)^{2}},
\label{eq:spiral1}
\end{equation}

\begin{equation}
R_{2}=\frac{1}{k} \sqrt{\left(\varphi + C_{2}+\pi\right)^{2}-(k z)^{2}}.
\label{eq:spiral2}
\end{equation}
where $m=$1 is for the first order vortex and $C_2$ is a constant. 

%-------------------------------------
\section{\label{sec:Appendix C} Role played by the aperture}
\begin{figure*} [!htbp]
\includegraphics[width=17cm]{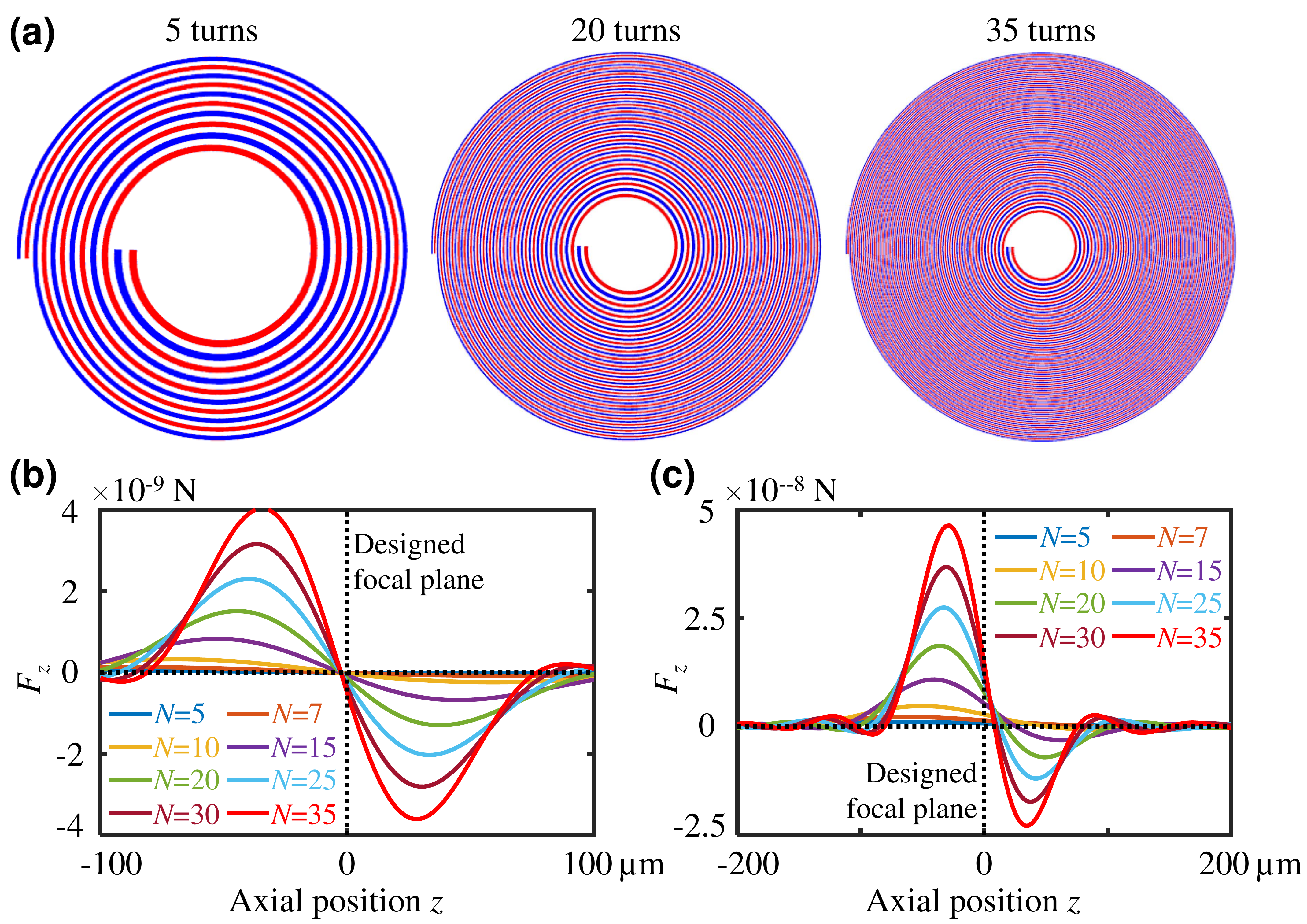}
\caption{ (a) Geometric shapes of spiral electrodes (IDTs) with different numbers of turns.
The axial radiation force $F_z$ versus different position $z$ for (b) a cell with radius $a=$ 11 $\mu$m and (c) a Pyrex sphere with $a=$ 9 $\mu$m, respectively, with a focused-vortex base acoustical tweezers at frequency $f_0=$ 40 MHz with different numbers of spiral turns. The scattering force contribution is relatively strong for a Pyrex sphere (elastic model) than for a cell (liquid model). The axial trapping positions for the cell are almost the same, while those for the Pyrex moves closer to the designed focal position as the number of spiral turn increases (as shown in the insert at the bottom-right corner), leading to the increase of focused ability (hence, gradient force).
The gravity effect could be ignored compared to the trapping force.}
\label{Fig7: Different turns.}
\end{figure*}

\begin{table*}[!htbp]
\small
  \caption{ Number of spiral turn $N$ for acoustical tweezers with same focal length 1 mm and related aperture angles}
  \label{table2}
  \begin{tabular*}{1\textwidth}{@{\extracolsep{\fill}} lcccccccc }
    \hline 
    $N$                        & 5    & 7     & 10   & 15   & 20   & 25   & 30   & 35   \\
    \hline 
    Aperture radius (mm)       & 0.74 & 0.86  & 1.02 & 1.27 & 1.50 & 1.72 & 1.94 & 2.15 \\
    Aperture angle (degrees)   & 36   & 41    & 46   & 52   & 56   & 60   & 63   & 65   \\
    \hline
  \end{tabular*}
\end{table*}

In the expressions of the two spiral electrodes given in Eqs. (\ref{eq:spiral1}) and (\ref{eq:spiral2}), the minimum and maximum azimuthal angle $\varphi_{min}$ and $\varphi_{max}$, which determine the aperture angle of the HAT can be chosen arbitrarily. For manipulation in microscopes, the minimum angle can be chosen to leave some space for observation near the central axis since generally the electrodes are not transparent
\cite{baudoin2019folding}. The maximum azimuthal angle proportional to the number of spiral turns plays an important role on the axial and lateral trapping capabilities. Here, lateral and axial trapping forces for different number of spiral turns with the corresponding aperture radii and angles given in TABLE \ref{table2} and three selected patterns in Fig. \ref{Fig7: Different turns.}(a). The axial radiation force versus the axial position $z$ for a cell with radius $a=$ 11 $\mu$m is given in (b), and for a Pyrex sphere with $a=$ 9 $\mu$m in (c). The axial trapping force increases with the increase of the aperture angle. For all the number of turns as discussed, the cell can be trapped in 3D, while for the Pyrex sphere, the axial trap is lost for the case with 5 turns. Indeed, the acoustic impedance of cells is smaller than that of Pyrex, leading to weaker scattering contribution which will push the cell/particle outside of the trap center. In addition, the axial trapping position $z_{trap}$ moves closer to the designed focal plane when the number of turns increases. This is because the restoring gradient force is enhanced more than the pushing scattering force.

\renewcommand\refname{Reference}
\bibliography{main}        %Produce the bibliography

\end{document}